\documentclass[10pt, conference, compsocconf]{IEEEtran}
\usepackage{amssymb,amsmath,epsfig}
\usepackage{graphicx,color}
\usepackage{cite}
\usepackage{multicol} 
\usepackage[keeplastbox]{flushend}

\begin{document}

\title{Multisensor Data Fusion for Water Quality Monitoring using Wireless Sensor Networks}

\author{\IEEEauthorblockN{Ebrahim~Karami,~\IEEEmembership{Member,~IEEE,}
        Francis M. ~Bui,~\IEEEmembership{Member,~IEEE,}
        and~Ha~H.~Nguyen~\IEEEmembership{Senior~Member,~IEEE} }
\IEEEauthorblockA{Department of Electrical and Computer Engineering, University of Saskatchewan, Canada}
\IEEEauthorblockA{email: \{ebk855,francis.bui,ha.nguyen\}@mun.ca}}

\maketitle

\begin{abstract}
In this paper, the application of hierarchical wireless sensor networks in water quality monitoring is investigated. Adopting a hierarchical structure, the set of sensors is divided into multiple clusters where the value of the sensing parameter is almost constant in each cluster. The members of each cluster transmit their sensing information to the local fusion center (LFC) of their corresponding cluster, where using some fusion
rule, the received information is combined, and then possibly sent to a higher-level central fusion center (CFC). A two-phase processing scheme is also envisioned, in which the first phase is dedicated to detection in the LFC, and the second phase is dedicated to estimation in both the LFC and the CFC. The focus of the present paper is on the problem of decision fusion at the LFC: we propose hard- and soft-decision maximum a posteriori (MAP) algorithms, which exhibit flexibility in minimizing the total cost imposed by incorrect detections in the first phase. The proposed algorithms are simulated and compared with conventional fusion techniques. It is shown that the proposed techniques result in lower cost. Furthermore, when the number of sensors or the amount of contamination increases, the performance gap between the proposed algorithms and the existing methods also widens.

\end{abstract}


\let\thefootnote\relax\footnotetext{This work was supported by an NSERC engage grant and IBM Canada.}

\section{Introduction}
Providing a reliable supply of potable water is an important goal in today’s society. To this end, water contamination warning systems (WCWSs) are typically deployed to monitor the quality of water. At the same time, wireless sensor networks (WSNs) have found extensive applications in monitoring physical or environmental conditions such as temperature, sound, pressure, etc. Therefore in this context, WCWSs have been one of the most recent embodiments of WSNs  \cite{Stoianov2003, Smith1997, Lin2004, Koch2011}. \\
 
 In a WCWS, the type of the parameters that must be monitored and controlled depends on the use of water. For example, for drinking water, chemical contamination is much more important to monitor; while for industrial applications, physical contaminations are more important, because physical objects in the water may damage industrial equipments that work with water \cite{Byer2005}. Therefore, when used in isolation, a
particular sensor may produce large sensing errors, responsible for incorrect or missing alarms. On the other hand, reliable and high-quality sensors, with lower variance in the sensing error, are invariably more expensive. As such, in a WCWS where many sensors may be needed to provide sufficient coverage of large geographical areas, e.g., a river or a water distribution system, the total cost may prove economically infeasible. \\
In addition, for a WSN, sensing error is not the only source of error: the quality of the wireless links is another major limiting factor. Therefore, to combat both sources of error, collaboration among the sensors in the network is useful in enabling distributed parameter estimation (DPE) \cite{Lin2004}. In DPE, each sensor is allowed to either send its measurement to a fusion center, or share it with other sensors which are in its
transmission range \cite{Ribeiro2006}.\\
The hierarchical structure increases the efficiency of the WSN in many aspects. In a hierarchical structure, the monitoring area is divided into multiple clusters where the value of the sensing parameter is almost constant inside each cluster but it may vary from one cluster to another. In general, the cross correlation between the values of the parameter in two clusters depends on some factors such as relative distance between the clusters, direction and velocity of the water flow. As shown in Fig. 1, each cluster has a local fusion center (LFC) which makes a local decision and then sends it to a central fusion center (CFC) to make the final decision. When a WSN is used to monitor the water quality, the sensing information from different types of sensors such as pH, DO (dissolved Oxygen), arsenic, permanganate and so on, are sensed and transmitted to the LFC \cite{Lin2004}. Using the hierarchical structure not only reduces the required power for communication between sensors and consequently increases the lifetime of the batteries, but also reduces the complexity of routing and scheduling in the wireless network. The cluster size can be optimized to minimize the required communications overhead among sensors \cite{Karami2011a}.\\

To increase the bandwidth efficiency in a DPE based system, one can perform monitoring in two phases. In the first phase, the quality of the water is constantly  monitored by sensors and each sensor periodically sends a binary signal to the LFC where 1 means that the water has become contaminated, and 0 means the water is healthy. Then the LFC combines the received binary signals using some fusion rule and if its final decision is whether the water in that area is contaminated, the system goes to the second phase where the LFC dedicates a much larger bandwidth to its members and asks them to send the values of the monitoring parameters for more exact processing and forwarding them to the CFC. This two-stage approach reduces the required bandwidth because the 
system mostly works in the first phase where only low-rate binary decisions are transmitted from sensors to the LFC and this communication needs much less bandwidth compared to the case where
the measured signal is directly sent to LFC. This paper focuses on the local detection in the LFC, i.e., phase 1, while deferring phase 2 to a future work.\\
The objective is to formulate an optimum detection for each LFC. Conventional fusion techniques for binary decisions are OR-rule, AND-rule, and $n$-out-of-$M$-rule. For the simpler OR-rule and AND-rule, the received binary decisions from the sensors are simply fed into a logical OR or AND operator to make the
final decision. However, with a more general $n$-out-of-$M$-rule, the final decision is a logical true, i.e., bit 1, if at least $n$ out of $M$ sensors have triggered alarm. Obviously the AND-rule results in a high rate of missed detections, making it unsuitable in WCWS where any contamination misdetection might be dangerous. On the other hand, the OR-rule leads to a high false alarm rate, so that the system may operate mostly in the second phase, where bandwidth and the other network resources are wasted. The $n$-out-of-$M$-rule offers a more flexible compromise, where the trade-off between missed detection and false alarm rates may be more properly balanced, based on a particular application scenario.\\

In applying fusion rules, a consideration that warrants attention is the utilization of sensor statistics. While conventional fusion techniques, including $n$-out-of-$M$-rule, do not utilize the sensor statistics directly, it is evident that a detection rule needs to exploit the available statistical information in order
to achieve optimality, especially when applied in a real implementation. To a certain extent, more recent works take into account the sensor statistics in optimizing the parameter $n$ \cite{Han2010}, \cite{Atapattu2011}. Nevertheless, the fact that the sensors may have different accuracy is not directly accounted for in this optimization. Furthermore, given that the parameter statistics may vary in a realistic situation, the value of $n$ would have to be accordingly adapted for optimality.\\

Other more statistically deriven fusion methods include the maximum likelihood (ML) and maximum a posteriori (MAP) algorithms. The ML detection minimizes the sum of false alarm and missed detection rates; as such, it outperforms the optimized $n$-out-of-$M$-rule \cite{Li2010}. Similarly, the MAP algorithm minimizes the total probability of error. However, it should be noted that, for practical water monitoring applications, the
probability of missed detection is typically more important than the probability of false alarm—a fact ignored by both the conventional ML and MAP algorithms.\\

In light of the above limitations of conventional data fusion in the context of water monitoring, this paper proposes a flexible MAP detector— that is nonetheless of low complexity— which minimizes the total cost imposed by false alarms and missed detections. In other words, by appropriately weighting these probabilities, a higher priority can be assigned to either quantity, to suit a particular application scenario; the proposed
MAP detector then minimizes this weighted error rate.

The rest of this paper is organized as follows. The system model and problem definition are presented in Sec. II. Fusion techniques based on the MAP algorithm, and their modifications are presented in Sec. III. 
Lastly, simulation results are presented in Sec. IV, with the conclusion in the Sec. V.


\section{System Model and Problem Definition}

\label{sec:system_model}
Consider a network with $N$ total sensors, used to monitor the quality of water in a measurement area, e.g., a pool or a river. As discussed in the above, this area is hierarchically divided into multiple clusters, as illustrated in Fig. 1. However, since the present paper addresses phase 1 of the processing scheme, i.e., fusion in the LFC, the focus is on a single cluster, with M active sensors. Let $\theta$ denote the parameter to be measured, which, by cluster selection, should be nearly constant at all $M$ sensors in the cluster. As noted previously, depending on the application, $\theta$ may represent quantities such as PH, DO, arsenic concentration, etc. Therefore, $\theta$ is generally a (bounded) continuous-valued quantity. Next, let $e_m$ be the sensing error at the $m$th sensor. Then, the measured parameter can be modeled as
\begin{equation}
\label{eqn:nonlinear_sensing}
\hat{\theta}_{m}=g(\theta,e_{m}),
\end{equation}
\noindent where $g(.)$ is a generalized function representing the sensor characteristic. With small errors in the sensor dynamic range, linear approximation provides the simplification 
 
\begin{equation}
\label{eqn:sensing}
\hat{\theta}_{m}=\alpha_{m}\theta+e_{m},
\end{equation}
\noindent where $\alpha_{m}=1$ if the sensors are properly calibrated.

\begin{figure}[tb]
	\centering
		\includegraphics[width=\columnwidth]{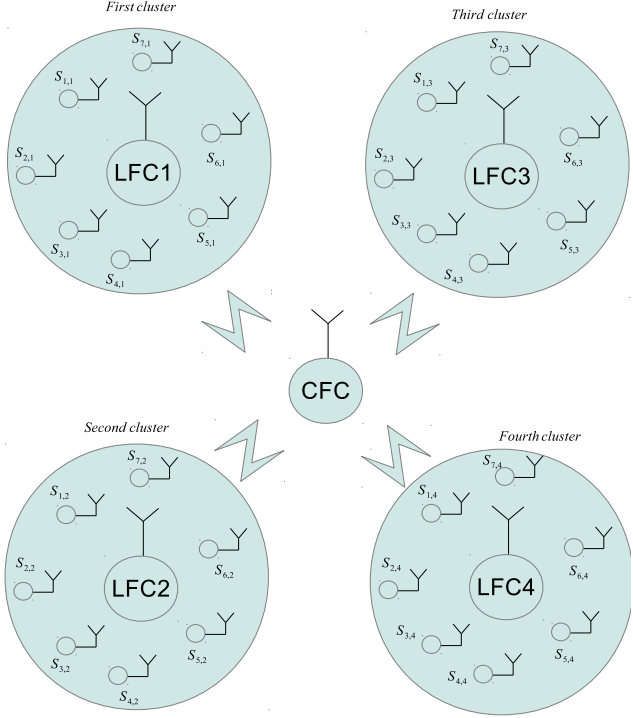}
	\caption{Distributed parameter estimation using a hierarchical WSN.}
	\label{fig:hier}
\end{figure}

The measured signal $\hat{\theta}_{m}$ is compared with $\tau_{max}$ and $\tau_{min}$  to decide whether to send decision signal $x_{m}=0$ or $1$ to the LFC.
In other words, if the measured pH is outside the safety range then a bit $1$ is modulated and sent to the LFC, otherwise bit $0$ is modulated and sent.
In the next step, decisions made by the members of the cluster are transmitted to the LFC through orthogonal channels. 
Since the transmitted signal is very narrow band, we can assume flat fading channel model. Assuming $h_{m}$ as the Rayleigh flat fading channel for the link between $m$th sensor and the LFC, $n_{m}$ 
as its corresponding additive complex Gaussian noise with variance $\sigma_n^2$, and binary phased shift keying modulation (BPSK) scheme, the received complex signal ${r}_{m}$ at the LFC is

\begin{equation}
\label{eq:rec_FC}
r_{m}=h_{m}(1-2x_{m})+n_{m}.
\end{equation}

The optimum fusion method is defined as finding the most probable estimation of $x$, for given set of observations $r_{m}$s where $x=0$ means water is safe and $x=1$ means water is contaminated. 
An optimum fusion rule can be defined for either soft detector or hard detector i.e., where $\hat{x}_{m}$s as hard estimations of the $x_{m}$s are used.
For the $m$th sensor, based on the precision of the sensor, two new parameters, $P_{F}^{m}$ as the probability of the false alarm and $P_{M}^{m}$ as the probability of missed detection, are defined at sensor level.
$P_{F}^{m}$ means the probability that the measuring parameter is inside its allowed range but the sensor detects it as outside the range. Likewise, $P_{M}$ means the probability that, the measuring parameter is outside its allowed range i.e. water is contaminated but the sensor does not detect the contamination. Probability of false alarm and missed detection for pH sensors are calculated from equations (\ref{eqn:dist_theta}) and (\ref{eqn:dist_error}) and using following equations,

\begin{equation}
\label{eqn:PF}
	P_{F}^{m}=P\left(x_{m}=1 |x=0\right).
\end{equation}
\begin{equation}
\label{eqn:PM}
	P_{M}^{m}=P(x_{m}=0 |x=1).
\end{equation}

With the distribution of the sensing parameter and sensing error and using (\ref{eqn:sensing}), (\ref{eqn:PF}), and(\ref{eqn:PM}), $P_{F}^{m}$ and $P_{M}^{m}$ can be easily calculated.

Since the wireless channel is noisy, the transmitted information is detected with some error. One can define the equivalent probabilities of missed detection and false alarm for each sensor as follows,

\begin{equation}
\label{eqn:PF_eq}
\tilde{P}_{F}^{m}=P_{F}^{m}(1-P_{b}^{m})+(1-P_{F}^{m})P_{b}^{m}.
\end{equation}

\begin{equation}
\label{eqn:PM_eq}
\tilde{P}_{M}^{m}=P_{M}^{m}(1-P_{b}^{m})+(1-P_{M}^{m})P_{b}^{m}.
\end{equation}

where $\tilde{P}_{F}^{m}$ and $\tilde{P}_{M}^{m}$ are the equivalent probabilities of false alarm and missed detection for the $m$th sensor respectively and $P_{b}^{m}$ is the bit error rate for signal received from the $m$th sensor, which depends on the quality of the channel from the sensor to the LFC.

$P_{F}^{m}$ and $P_{M}^{m}$ are considered as two basic criteria to compare the performance of the sensing techniques at either the sensor level or the LFC level, i.e., for the final decision $x^{tot}$ made by the LFC. In this case, we call them total probabilities of false alarm $P_{F}^{tot}$ and total probability of missed detection $P_{M}^{tot}$ defined as follow,
\begin{equation}
\label{eqn:PFt}
	P_{F}^{tot}=P\left(x^{tot}=1 |x=0\right),
\end{equation}
\begin{equation}
\label{eqn:PMt}
	P_{M}^{tot}=P(x^{tot}=0 |x=1).
\end{equation}

The total probability of error, $P_{E}^{tot}$, is calculated as follows:

\begin{equation}
\label{eqn:err0}
P_{E}^{tot}=P\left( x^{tot}=1,x=0\right)+P\left( x^{tot}=0,x=1\right).
\end{equation}

By substituting (\ref{eqn:PFt}) and (\ref{eqn:PMt}) in (\ref{eqn:err0}) , we have

\begin{equation}
\label{eqn:err}
P_{E}^{tot}=P_{H_{1}}P_{M}^{tot}+P_{H_{0}}P_{F}^{tot},
\end{equation}

where $P_{H_{0}}=P(x=0)$ and $P_{H_{1}}=P(x=1)$.

\section{MAP Fusion Technique}
This section presents both hard detection (HD) and soft detection (SD) based MAP fusion rules.
\subsection{SD MAP Fusion Algorithm}

The SD MAP fusion algorithm is defined as 
\begin{equation}
\label{eq:MAPs1}
x^{SD-MAP}=\text{arg} \hspace{5 mm} \text{max}_{x} \hspace{5mm} P(x|r_{1},r_{2},...,r_{M-1},r_{M}).
\end{equation}

(\ref{eq:MAPs1}) is conventionally solved by likelihood ratio (LR) test \cite{Vtrees}. Therefore likelihood ratio $\zeta^{SD}$ is defined as follows,

\begin{equation}
\label{eq:MAPs2}
\zeta^{SD}=\frac{P\left(x^{tot}=1\right)}{P\left(x^{tot}=0\right)},
\end{equation}

and using Bayes'rule, (\ref{eq:MAPs2}) is expanded as

\begin{equation}
\label{eq:MAPs20}
\zeta^{SD}= \prod_{m=1}^{M} \frac{P(r_{m}|x=1)}{P(r_{m}|x=0)}\frac{P(x=1)}{P(x=0)},
\end{equation}

and then using (\ref{eq:rec_FC}), (\ref{eq:MAPs20}) is calculated by (\ref{eq:MAPs3})

\begin{figure*} 
\begin{equation}
\label{eq:MAPs3}
\zeta^{SD}= \prod_{m=1}^{M} \frac{P(r_{m}|x_{m}=1)P(x_{m}=1|x=1)+P(r_{m}|x_{m}=0)P(x_{m}=0|x=1)}{P(r_{m}|x_{m}=1)P(x_{m}=1|x=0)+P(r_{m}|x_{m}=0)P(x_{m}=0|x=0)}\frac{P(x=1)}{P(x=0)},
\end{equation}
\end{figure*}

and from (\ref{eq:rec_FC}), (\ref{eqn:PF}), and (\ref{eqn:PM}) $\zeta^{SD}$ is calculated by (\ref{eq:MAPs4}) 
\begin{figure*}
\begin{equation}
\label{eq:MAPs4}
\zeta^{SD}= \prod_{m=1}^{M} \frac{\exp(-\frac{|r_m+h_m|^2}{\sigma_{n}^{2}})(1-P_{M}^{m})+\exp(-\frac{|r_m-h_m|^2}{\sigma_{n}^{2}})P_{M}^{m}}{
\exp(-\frac{|r_m+h_m|^2}{\sigma_{n}^{2}})P_{F}^{m}+\exp(-\frac{|r_m-h_m|^2}{\sigma_{n}^{2}})(1-P_{F}^{m})}\frac{P(x=1)}{P(x=0)},
\end{equation}
\end{figure*}

Consequently if $\zeta^{SD} > 1$ then $x^{SD-MAP}=1$ and otherwise $x^{SD-MAP}=0$.

\subsection{HD MAP Fusion Algorithm} 
\label{sec:MAP}

The HD MAP  fusion algorithm is defined as

\begin{equation}
\label{eq:MAP1}
x^{HD}=\text{arg} \hspace{5 mm} \text{max}_{x} \hspace{5mm} P(x|\hat{x}_{1},\hat{x}_{2},...,\hat{x}_{M-1},\hat{x}_{M})
\end{equation}

And consequently its LR is defined as,

\begin{equation}
\label{eq:MAP2}
\zeta^{HD}=\frac{P\left(x^{tot}=1\right)}{P\left(x^{tot}=0\right)}= \prod_{m=1}^{M} \frac{P(\hat{x}_{m}|x=1)}{P(\hat{x}_{m}|x=0)}\frac{P(x=1)}{P(x=0)}
\end{equation}

From (\ref{eqn:PF_eq}) and (\ref{eqn:PM_eq}), we can see in the right hand side of (\ref{eq:MAP2}) if $\hat{x}_{m}=1$ then $P(\hat{x}_{m}|x=1)=1-\tilde{P}_{M}^{m}$ and $P(\hat{x}_{m}|x=0)=\tilde{P}_{F}^{m}$
 and on the other hand if $\hat{x}_{m}=0$ then $P(\hat{x}_{m}|x=1)=\tilde{P}_{M}^{m}$ and $P(\hat{x}_{m}|x=0)=1-\tilde{P}_{F}^{m}$. consequently
\begin{equation}
\label{eq:MAP3}
P \left(\hat{x}_{m}|x=1 \right)=\hat{x}_{m}\left(1-\tilde{P}_{M}^{m}\right)+\left(1-\hat{x}_{m}\right)\tilde{P}_{M}^{m},
\end{equation}

\begin{equation}
\label{eq:MAP4}
P\left(\hat{x}_{m}|x=0\right)=\hat{x}_{m}\tilde{P}_{F}^{m}+\left(1-\hat{x}_{m}\right)\left(1-\tilde{P}_{F}^{m}\right),
\end{equation}

and therefore (\ref{eq:MAP2}) is calculated as

\begin{equation}
\label{eq:MAP}
\zeta^{HD}= \prod_{m=1}^{M}\left( \beta_{m}\hat{x}_{m}+\ (1-\hat{x}_{m}) \alpha_{m}\right)  \frac{P(x=1)}{P(x=0)}.
\end{equation}
where $\beta_{m} =\frac{1-\tilde{P}_{M}^{m}}{\tilde{P}_{F}^{m}}$ and $\alpha_{m}=\frac{\tilde{P}_{M}^{m}}{1-\tilde{P}_{F}^{m}}$ and like HD MAP, if $\zeta^{HD} > 1$ then $x^{SD-MAP}=1$ and otherwise $x^{SD-MAP}=0$. Right hand side of (\ref{eq:MAP}) shows that MAP fusion algorithm
 can be interpreted as kind of weighted $n$-out-of-$M$-rule where weights and the value of $n$ has been optimized! To prove this similarity assume an special case where all sensors have
 the same quality and the channel link condition which result the same $\tilde{P}_{F}^{m}$ and $\tilde{P}_{M}^{m}$ for them. In this case if we assume $n$ as number of ones and $M-n$
 as number of zeros in detected decisions, then we have
\begin{equation}
\label{eq:MAP_simple1}
\zeta^{HD}= \left(\frac{1-\tilde{P}_{M}}{\tilde{P}_{F}}\right)^{n}\left(\frac{\tilde{P}_{M}}{1-\tilde{P}_{F}}\right)^{M-n}\frac{P(x=1)}{P(x=0)}.
\end{equation}

 if we define $\alpha=\frac{\tilde{P}_{M}}{1-\tilde{P}_{F}}$, $\beta=\frac{\tilde{P}_{M}}{1-\tilde{P}_{F}}$ and $\gamma=\frac{P\left(x=1\right)}{P\left(x=0\right)}\beta^{-M}$, then (\ref{eq:MAP_simple1}) is simplified as

\begin{equation}
\label{eq:MAP_simple2}
\zeta^{HD}= \left(\frac{\alpha}{\beta}\right)^{n}=\gamma.
\end{equation}

and obviously $\zeta^{HD} > 1$ if and only if $n \ge \frac{\log\left(\gamma\right)}{\log\left(\alpha\right)\log\left(\beta\right)}$. 
But if sensors do not have the same quality, MAP fusion performs the same as weighted $n$-out-of-$M$-rule where detected signals from each sensors multiplied with some weights which is dependents to its sensing quality, link quality and their value.

\subsection{Risk Management Using Modified MAP algorithm}
Either SD or HD MAP algorithms minimize $P_E^{tot}$ defined in (\ref{eqn:err}). In other word, in the definition of the $P_E^{tot}$, the importance of one false alarm is exactly the same as one missed detection.
But as we know, in the water monitoring the extra cost imposed by one missed detection is much more than one false alarm. In a two-phase WCWS, a false alarm pushes the system from the first phase to the second phase where more bandwidth, power and processing is required and therefore it imposes extra cost for these items.
On the other hand, a missed detection causes more severe damages because it directly affects health-based standards. Therefore we need to substitute $P_E^{tot}$ which is minimized by the conventional MAP algorithm with a new cost function as follows \cite{Barkat2005},

\begin{equation}
\label{eqn:errn}
\begin{split}
\bar{C}=C_{11}P_{H_{1}} (1-P_{M}^{tot})+C_{01} P_{H_{1}} P_{M}^{tot}\\
+C_{10}P_{H_{0}}P_{F}^{tot}+C_{00}P_{H_{0}}(1-P_{F}^{tot}),
\end{split}
\end{equation}

where $C_{ij}$ is the cost caused by hypothesis $H_{j}$ if $H_{i}$ is detected. The optimum fusion rule which minimizes (\ref{eqn:errn}) is achieved by modification of the LR ratio in either SD or HD MAP algorithm as follows,

\begin{equation}
\label{eqn:errn1}
\zeta^{Modified-HD}=\frac{C_{01}-C_{11}}{C_{10}-C_{11}}\zeta^{Modified-HD}.
\end{equation}

\section{Simulation Results}
In this Section, simulation results for the conventional and modified HD and SD MAP algorithms and their modified versions  is presented and compared with other fusion rules
such as MAX-rule and ML. 

\subsection{Case Study}

\begin{figure}[tb]
	\centering
		\includegraphics[width=\columnwidth]{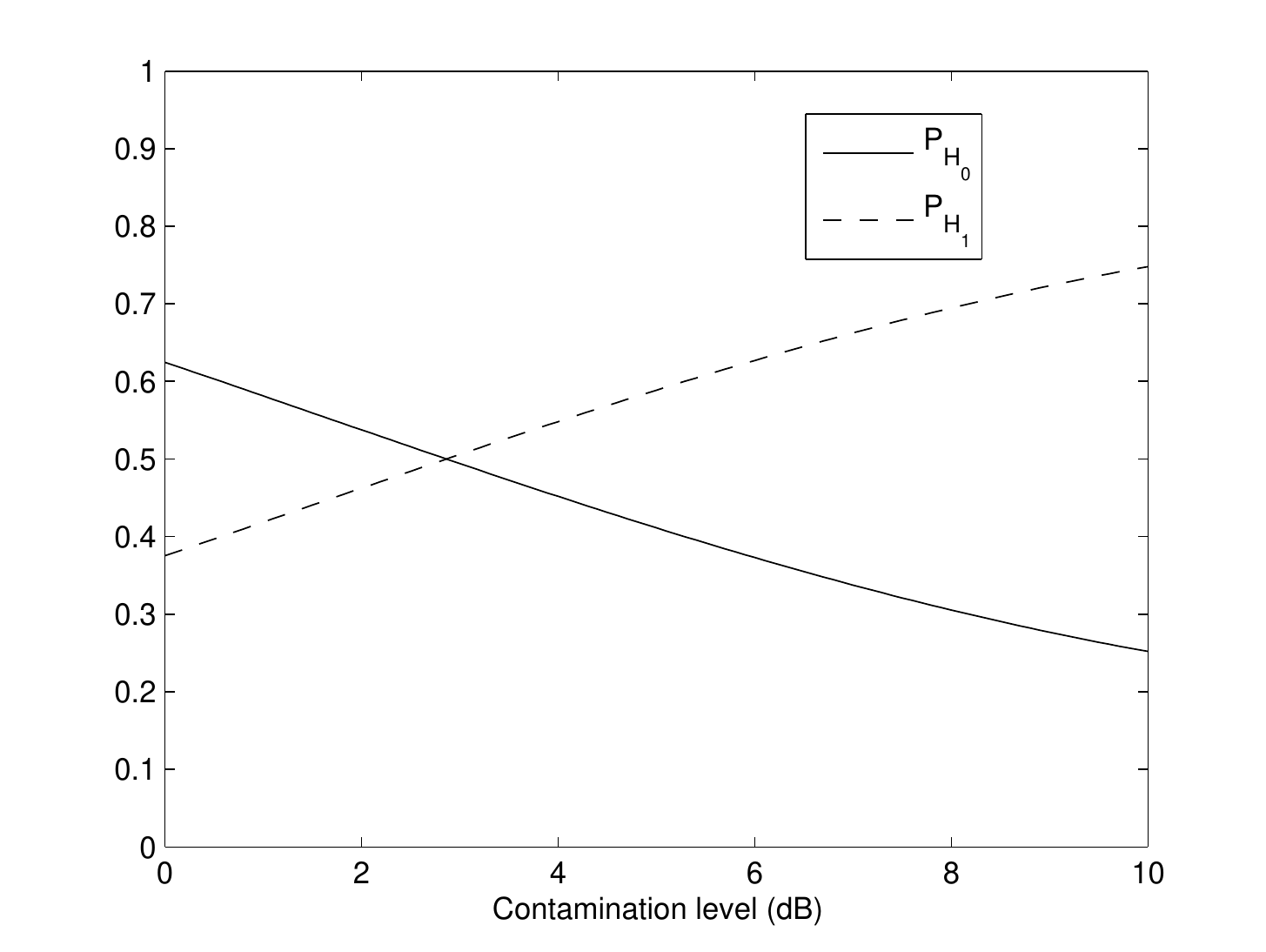}
	\caption{$P_{H_{0}}$ and $P_{H_{1}}$ vs. contamination level.}
	\label{ProbH}
\end{figure}

\begin{figure}[tb]
	\centering
		\includegraphics[width=\columnwidth]{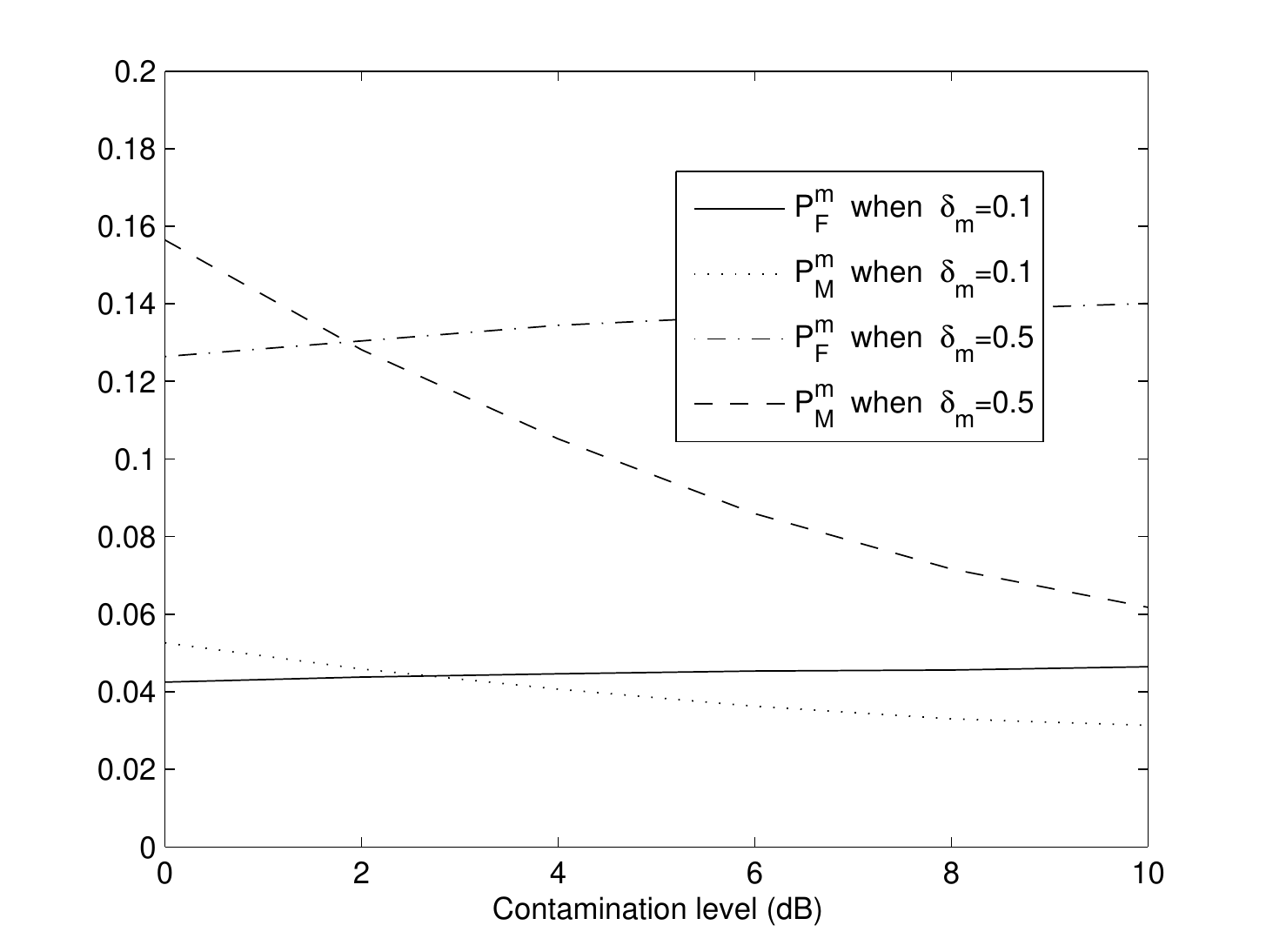}
	\caption{$P_{H_{0}}$ and $P_{H_{1}}$ vs. contamination level.}
	\label{PF_PM}
\end{figure}

In this paper we consider concatenated Gaussian distribution for the pH parameter and a uniform distribution for sensing error as follows,

\begin{equation}
\label{eqn:dist_theta}
P(\theta)=\frac{\exp\left(-\frac{(\theta-\theta_0)^2}{2\sigma^2}\right)}{\sqrt{2\pi}\sigma\left[F(\frac{\theta_{max}-\theta_0}{\sigma})-F(\frac{\theta_{min}-\theta_0}{\sigma})\right]},
\end{equation}
\begin{equation}
\label{eqn:dist_error}
P(e_{m})=\frac{1}{2\delta_{m}}, \hspace{10 mm} \text{if} \hspace{5 mm}  -\delta_{m}<e_{m}< +\delta_{m},
\end{equation}
where $\theta_{max}$, $\theta_{min}$, and $\theta_0$ are, respectively, maximum, minimum, and middle value of the sensing parameter and for the pH parameter they are 14, 0, and 7 respectively and $\delta_{m}$ is a parameter indicating the maximum deviation of the measured signal from its actual value.
For the distributions assumed for sensing parameter and sensing error as (\ref{eqn:dist_theta}) and (\ref{eqn:dist_error}), respectively, $P_{H_{0}}$ and $P_{H_{1}}$ are calculated as follows,
\begin{equation}
\label{eqn:H0}
P(\theta)=\frac{F(\frac{\tau_{max}-\theta_0}{\sigma})-F(\frac{\tau_{min}-\theta_0}{\sigma})}{F(\frac{\theta_{max}-\theta_0}{\sigma})-F(\frac{\theta_{min}-\theta_0}{\sigma})},
\end{equation}

\begin{equation}
\label{eqn:H1}
P(\theta)=1-P_{H_{0}},
\end{equation}

where for the pH of the drinking water $\tau_{max}=8.5$ and $\tau_{min}=6.5$ which are limits for basicity and acidity of the water, respectively. Figure \ref{ProbH} show $P_{H_{0}}$ and $P_{H_{1}}$ versus the variance of the contamination. 
By substituting (\ref{eqn:dist_theta}) and (\ref{eqn:dist_error}) in (\ref{eqn:PF}) and (\ref{eqn:PM}), $P_{F}^{m}$ and $P_{M}^{m}$ are also calculated. 
Figure \ref{PF_PM} shows the changes of $P_{F}^{m}$ and $P_{M}^{m}$ versus the variance of the contamination level.

\subsection{Simulation Setup}

As we mentioned before, in a WCWS, $P_{M}^{tot}$ is much more important than $P_{F}^{tot}$ and therefore, we set $C_{11}=C_{00}=0$, $C_{10}=1$, and $C_{01}=10$. Channel links are randomly generated from Rayleigh distribution. Contamination level and sensing error distribution have distributions as (\ref{eqn:dist_theta}) and (\ref{eqn:dist_error}).
The value of sensing error limit $\delta_{m}$ for each sensor is randomly generated and their average in each cluster is assume to be 0.1, 0.2, and 0.5 i.e. in all cases $1 \ge \delta_{m} \ge 0$. 
 Monte Carlo technique is used for simulations and results are averaged over 100000 independent runs. Contamination level $\sigma_{m}^2$ varies from 0dB to 10dB and average SNR for each sensor
 to LFC is 10dB.  $\bar{C}$ is considered as performance criteria.

\subsection{Results}
\begin{figure}[tb]
	\centering
		\includegraphics[width=\columnwidth]{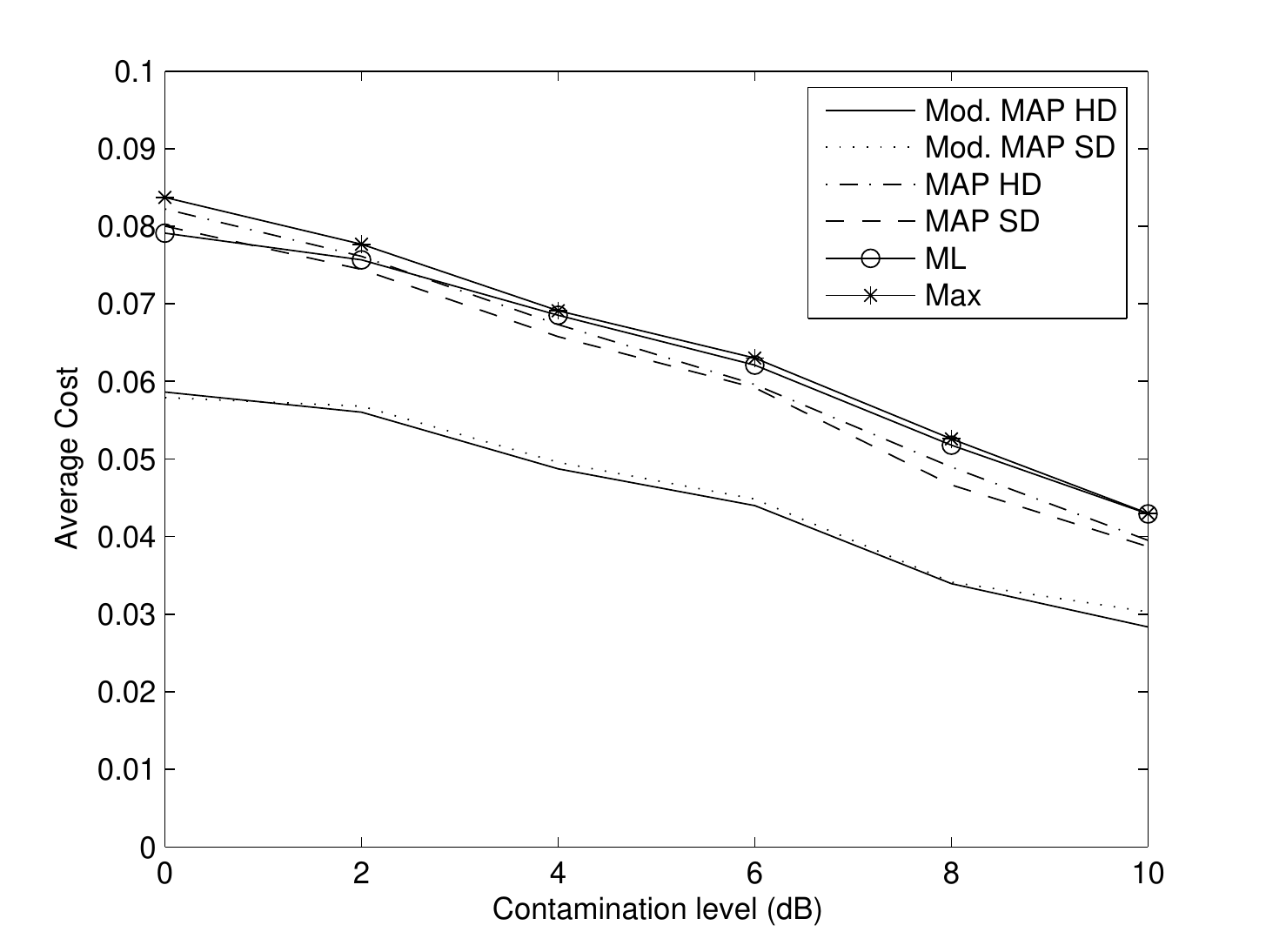}
	\caption{Average cost vs. contamination level for $M$=5 and $\bar{\delta}=0.1$.}
	\label{5s1t}
\end{figure}

\begin{figure}[tb]
	\centering
		\includegraphics[width=\columnwidth]{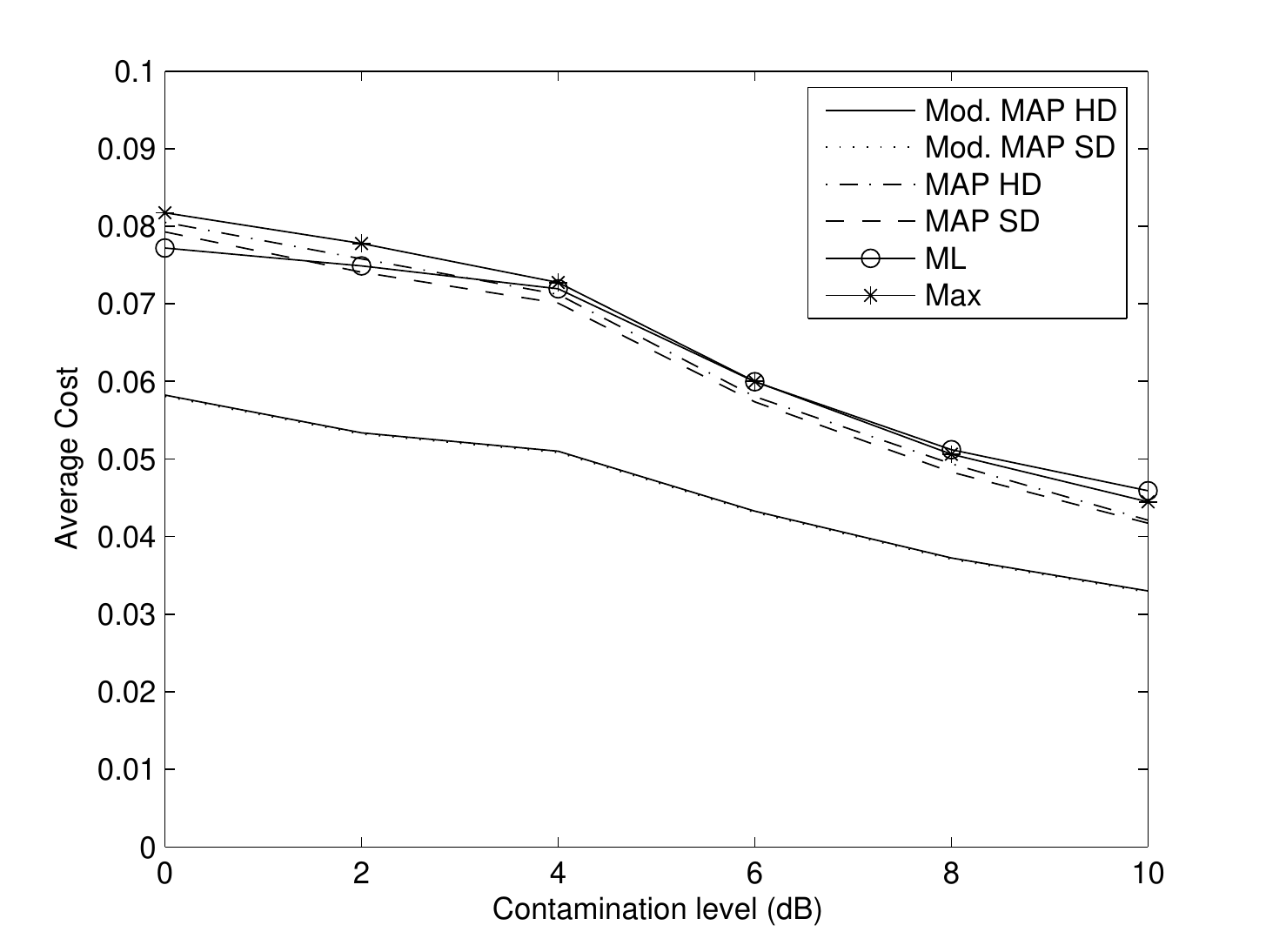}
	\caption{Average cost vs. contamination level for $M$=5 and $\bar{\delta}=0.2$.}
	\label{5s2t}
\end{figure}

\begin{figure}[tb]
	\centering
		\includegraphics[width=\columnwidth]{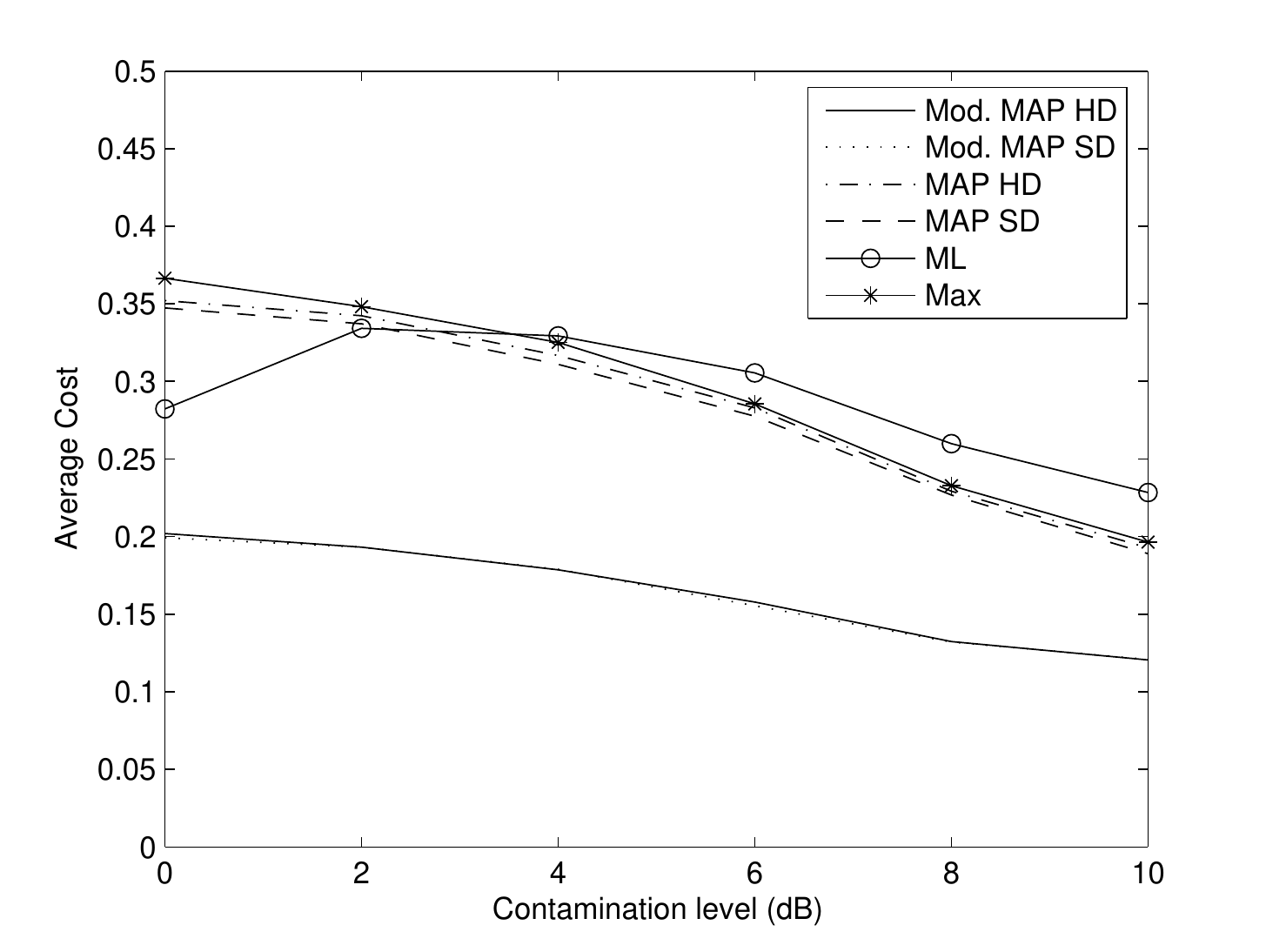}
	\caption{Average cost vs. contamination level for $M$=5 and $\bar{\delta}=0.5$.}
	\label{5s3t}
\end{figure}

\begin{figure}[tb]
	\centering
		\includegraphics[width=\columnwidth]{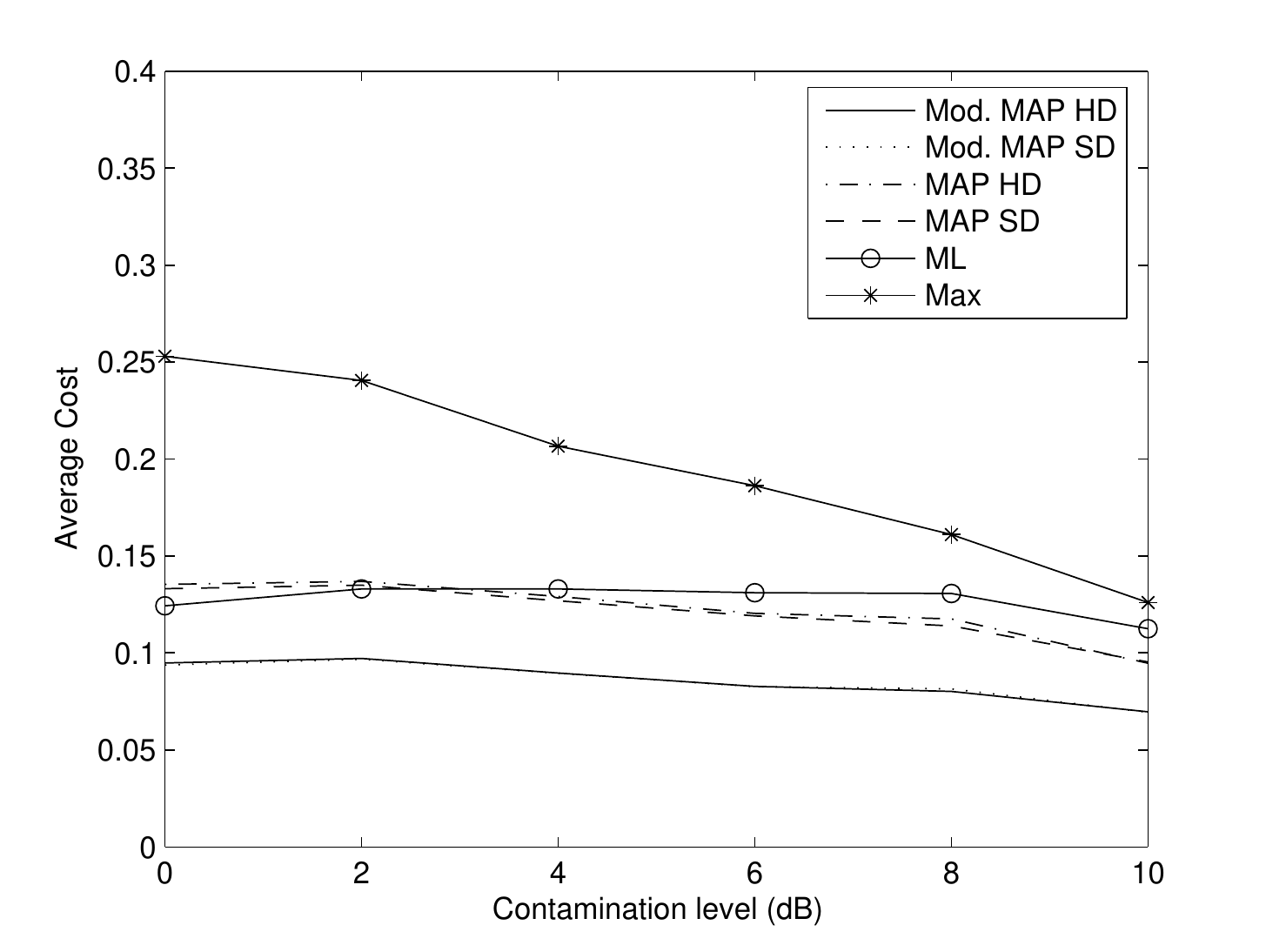}
	\caption{Average cost vs. contamination level for $M$=10 and $\bar{\delta}=0.5$.}
	\label{10s3t}
\end{figure}

\begin{figure}[tb]
	\centering
		\includegraphics[width=\columnwidth]{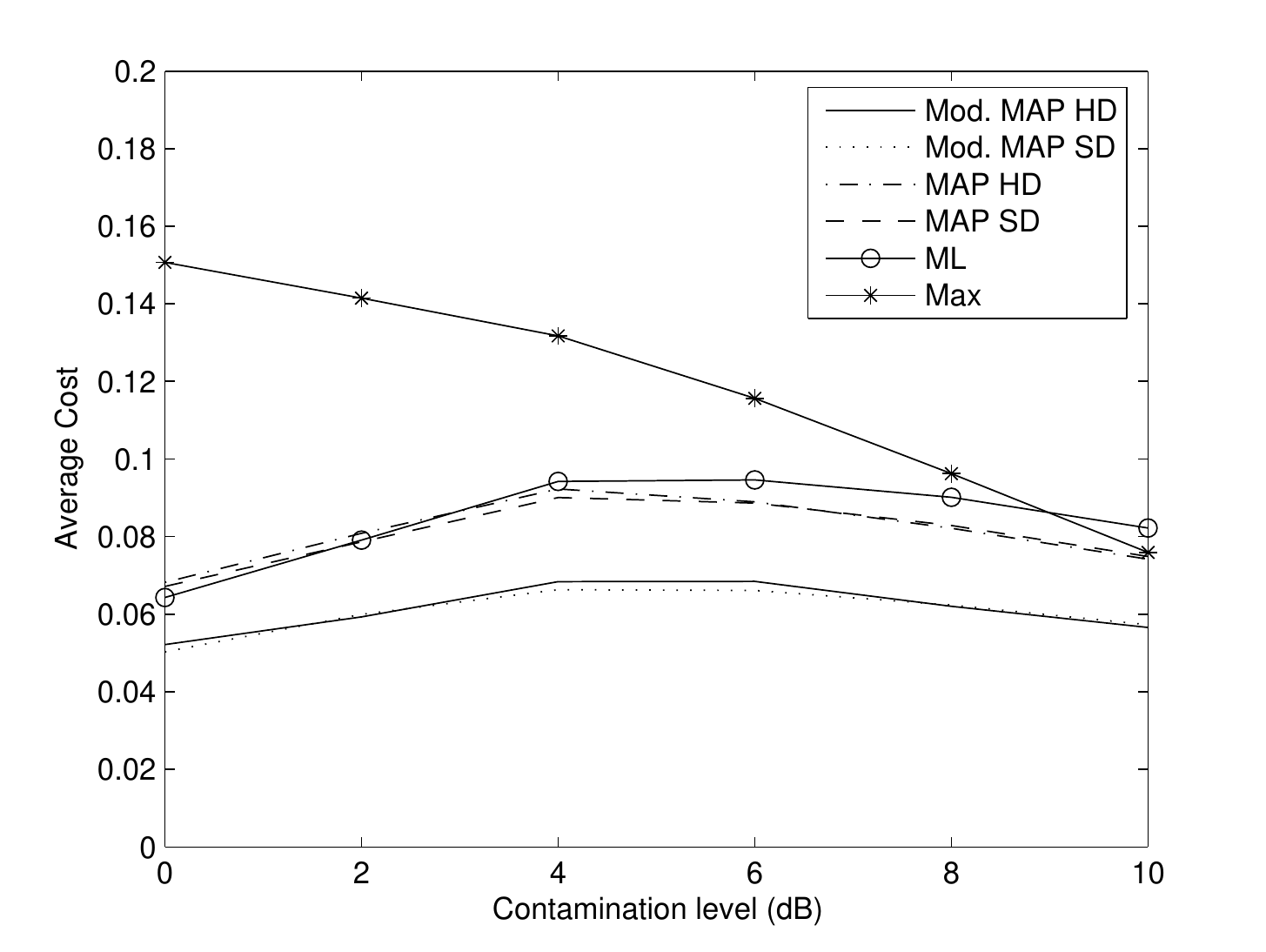}
	\caption{Average cost vs. contamination level for $M$=20 and $\bar{\delta}=0.5$..}
	\label{20s3t}
\end{figure}

 Figures \ref{5s1t}, \ref{5s2t}, and \ref{5s3t}  present $\bar{C}$ versus contamination level for $M=5$ and $\bar{\delta}=0.1$, $0.2$ and $0.5$.  In all cases we can see, the HD MAP algorithm performs very close to the SD MAP algorithm and also the Modified HD MAP algorithm performs very close 
 to the modified SD MAP modified. Consequently we can see that using the HD based algorithms which need much less computational complexity are more reasonable to use. In all cases, modified MAP algorithms outperform the other ones. On the other hand, in all cases conventional and in high values of contamination level, HD and SD MAP algorithms perform just a little better than ML but in lower amount of contamination ML presents lower cost because in this case $P_{H_{1}}$ is less than
 $P_{H_{0}}$ and consequently LR of the ML algorithm is closer to the LR of the modified MAP.In the most of the scenarios, the MAX-rule is the worst one. \\By comparing  Figures \ref{5s1t}, \ref{5s2t}, and \ref{5s3t} we can see that 
 as we expect, when $\bar{\delta}$ increases, the $\bar{C}$ increases too. In this case, in low contamination levels the gap between modified MAP algorithms and the other ones does not change. But in larger contaminations level this gap is more than Fig. 1 and this shows that in large values of contamination level, other algorithms are farer from optimality. 
Figures \ref{10s3t} and \ref{20s3t} present the simulation results for the worst sensor case i.e. when number of sensors is 10 and 20 sensors respectively. Except the last Figure i.e. when $\bar{\delta}=0.5$ and $M=20$ in other cases, $\bar{C}$ decreases with the increasing of the contamination level. This means that when the contamination level increases, the fusion rules have higher chance to detect the existence of the contamination correctly. 
But when $M=20$ of the lowest quality sensors is used, in low values of the contamination level, increasing the contamination does not improve the accuracy of the fusion rules. 

\section{Conclusion}
In this paper, we proposed maximum a posteriori (MAP) based fusion rules for the application of the wireless sensor networks in water contamination detection systems. 
Since the conventional MAP algorithms give the same value to the false alarms and missed detections and in the water contamination systems missed detections are much more important, we modified the conventional MAP to minimize a new cost function which 
pays higher penalty to the missed detection than false alarms. The proposed MAP and modified MAP algorithms were simulated and compared with conventional fusion rules and it was shown that the modified MAP algorithms  present much lower average cost.



\bibliographystyle{IEEEtran}


\end{document}